\documentclass[a4paper,10pt, twocolumn]{article}
\usepackage[utf8]{inputenc}
\usepackage[top=1.5cm, left=1.5cm, right=1.5cm, bottom=1.5cm]{geometry}

\usepackage{amsfonts,amssymb,amsmath,amsthm,dsfont}

\usepackage{subfigure}
\usepackage{color}

\usepackage{graphicx}

\makeatletter
\let\NAT@parse\undefined
\makeatother
\usepackage[sort&compress, numbers]{natbib}

\usepackage{algorithm,algorithmic}
\usepackage[ruled,vlined,algosection,algo2e]{algorithm2e} % algorithm environment
\graphicspath{{./},{images/}}
\usepackage{hyperref}

\usepackage[footnotesize]{caption}

\renewcommand{\leq}{\leqslant}
\renewcommand{\geq}{\geqslant}

\DeclareMathOperator*{\argmin}{argmin}

\newcommand{\bs}{\boldsymbol}

\newcommand{\ie}{\emph{i.e.}, }
\newcommand{\eg}{\emph{e.g.}, }
\newcommand{\etal}{\emph{et al.}}

\newcommand{\Lmodel}{\bar{\bs L}}
\newcommand{\Xmodel}{\bar{\bs x}}

\newcommand{\sq}{\vspace{-1mm}}

\usepackage{comment}

\makeatletter
\renewcommand{\section}{\@startsection {section}{1}{\z@}%
{-3.5ex \@plus -1ex \@minus -.2ex}%
{2.3ex \@plus.2ex}%
{\normalfont\large\bfseries}}
\makeatother 

\title{ \vspace{-0.5cm}Morphological components analysis for circumstellar disks imaging}
\author{Beno\^{i}t Pairet$^1$\thanks{BP and LJ are funded by the Belgian F.R.S.-FNRS.}, Faustine Cantalloube$^2$, Laurent Jacques$^{1*}$\\
\footnotesize $^1$ISPGroup, ICTEAM/ELEN, UCLouvain, Belgium\ $^2$ Max Planck Institute for Astronomy, Germany
}

\date{\empty}

\renewenvironment{abstract}{\noindent\bf\small {\em Abstract---}}{}

\begin{document}

\maketitle

\begin{abstract}
Recent developments in astronomical observations enable direct imaging of circumstellar disks. Precise characterization of such extended structure is essential to our understanding of stellar systems. However, the faint intensity of the circumstellar disks compared to the brightness of the host star compels astronomers to use tailored observation strategies, in addition to state-of-the-art optical devices. Even then, extracting the signal of circumstellar disks heavily relies on post-processing techniques. In this work, we propose a morphological component analysis (MCA) approach that leverages low-complexity models of both the disks and the stellar light corrupting the data. In addition to disks, our method allows to image exoplanets. Our approach is tested through numerical experiments.
\end{abstract}

\sq
\section{Introduction}
\sq

Direct imaging of stellar systems is a challenging task that requires hardware with high contrast and high resolution to be able to detect the faint objects located near the extremely bright star. Ground based telescope achieve the highest resolution and thanks to adaptive optics (AO), they are able to overcome the atmospheric turbulence. The brightness of the star is dimmed using a coronagraph. However, even with state-of-the-art hardware, residual quasi-statics non-common path aberrations, form \emph{speckles} in the observations~\cite{fitzgerald2006speckle}. The presence of these speckles prevent the detection of on-sky signals such as disks and exoplanets. Specific observation strategies and post-processing techniques have been used to force diversity within the data and leverage this diversity to improve the separability between speckles and on-sky signals.

Angular differential imaging (ADI) is a popular observation strategy that leverages the rotation of the Earth to introduce such a data diversity. In this setting, the star is kept in the center of the field of view while snapshots of the stellar systems are taken during the observation period~(\ie a few hours). As most of the residual speckles (after AO) are due to the telescope itself, they remain quasi-static in the images, while on-sky signals follow a deterministic circular trajectory, determined by known parallactic angles~\cite{marois2006angular}.

The typical ADI data processing pipeline is guided by a principal component analysis (PCA)~\cite{soummer2012detection,amara2012pynpoint}: \textit{(i)} the $T \times (n\times n)$ spatiotemporal data cube (with $T$ the number of frames and $n^2>T$ the number of pixels of each frame) is reshaped into a $\mathbb R^{T\times n^2}$ matrix $\bs Y$, \textit{(ii)}  its rank-$r$ approximation $\bs L$ is computed by thresholding its singular values decomposition (SVD) to its $r$ largest singular values, \textit{(iii)} $\bs L$ is then subtracted from the data to form $\bs S = \bs Y - \bs L$ containing the on-sky signals, and \textit{(iv)} the frames of $\bs S$ are aligned (by rotation and interpolation) to the on-sky coordinates and temporally averaged to form the processed frame. Objects detection can then be done by hypothesis testing~\cite{mawet2014fundamental,pairet2019stim}. 

\begin{figure}
  \centering
  {\includegraphics[width=0.23\textwidth]{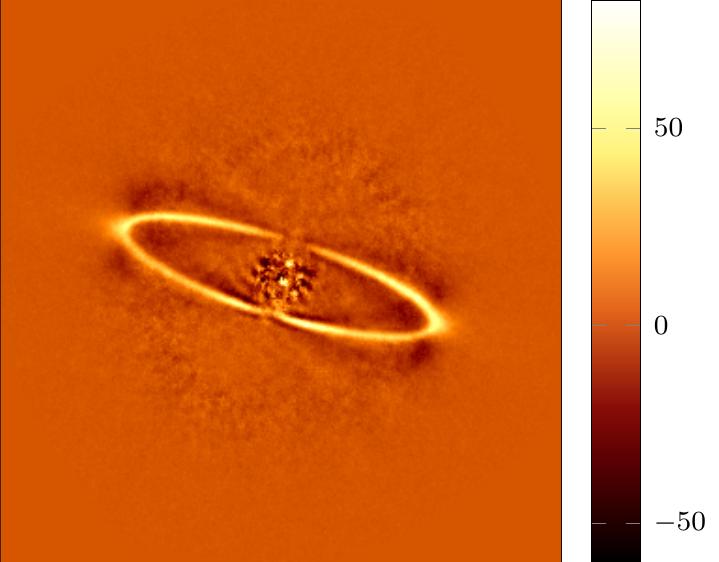}}
  {\includegraphics[width=0.23\textwidth]{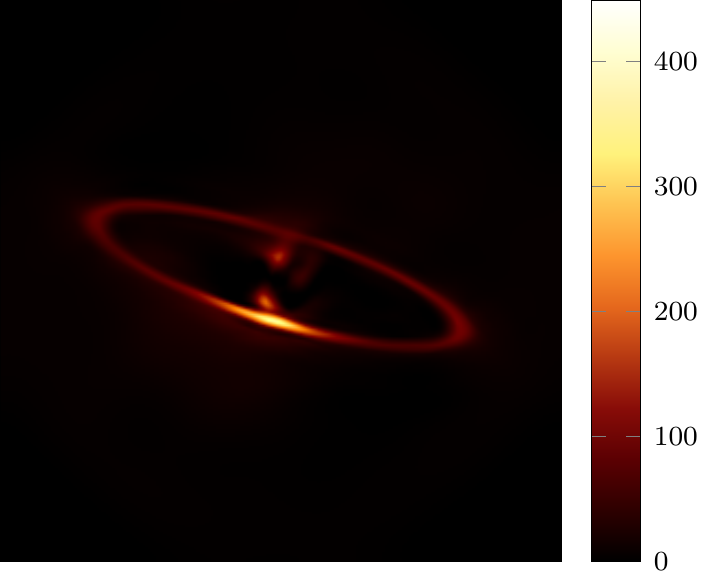}}
\caption{Processed frame for HR~4796A with PCA (left) and with our MCA approach (right).}
\label{fig:HR4796A}
\end{figure}

\textbf{Morphological Limitations:} The morphology of the circumstellar disks is known to be severely distorted by the PCA pipeline, hindering our capability to study disks structures from ADI datasets. Results of PCA on the ellipsoidal disk surrounding HR~4796A~\cite{milli2017near} is displayed on Fig.~\ref{fig:HR4796A} (left), where we can see unphysical artifacts with negative intensity. Among the few attempts to remove these artifacts, Milli~\etal~\cite{milli2012impact} reduced them by injecting forward-modeled disks. In~\cite{ren2018non}, non-negative matrix factorization was also used to reduce the artifacts. 

\sq
\section{Our approach}
\sq
We propose to recast the stellar system imaging task from ADI dataset as a MCA task~\cite{starck2005morphological,bobin2007morphological,donoho2009geometric}. To achieve this, we first present the acquisition model and then propose a constrained convex optimization solving our MCA problem. We then list the main physical priors our convex optimization is based upon.

\textbf{Acquisition model:} We restart from an ADI sequence $\bs Y \in \mathbb R^{T\times n^2}$, assumed to contain a disk. We model $\bs Y$ as the sum of two terms: the starlight $\bs Y_{*}$ and the rotating on-sky signal $\bs Y_{\odot}$. As the star is far from the Earth, it is point-like source and ideally, its intensity is blocked by the coronagraph. However, as discussed earlier, atmospheric turbulence and imperfections in the optics introduce speckles in the observation. These speckles are modeled as the sum of two terms, encoding their temporal behavior: a static term $\Lmodel$ and a non-static term $\bs N_s$: 
\begin{equation}
\textstyle \bs Y_{*} = \Lmodel+ \bs N_s
\end{equation}
where $\Lmodel$ is assumed to be a rank-$r$ matrix, $r\leq t$.

Concerning the on-sky component, its light intensity being small, we can neglect the non-common path aberrations. We thus assume its intensity to be constant through time and we model $\bs Y_{\odot}$ as a single rotating image $\Xmodel \in \mathbb R^{n^2}$
\begin{equation}
\textstyle \bs Y_{\odot} = R(\bs p \Xmodel^\top )
\end{equation}
where $\bs p \in \mathbb R^T$ stands for the time variation intensity of $\bs Y_{\odot}$ and $R : \mathbb R^{T\times n^2} \rightarrow \mathbb R^{T\times n^2} $ is the linear operator that rotates each frame of the volume according to the parallactic angles.

Furthermore, because the light is diffracted as it enters the telescope, the on-sky signal is blurred by the known telescope PSF $\varphi$. We write this as 
\begin{equation}
\textstyle \bs Y_{\odot} = \varphi * [ R\bs p (\Xmodel)^\top )],
\end{equation}
where $*$ denotes the 2D convolution applied separately on each image of $R(\bs p (\Xmodel)^\top $. 

The final acquisition model of $\bs Y$ reads
\begin{equation}
\textstyle\bs Y = \Lmodel +\bs N_s +  \varphi * [ R(\bs p \bs \Xmodel^\top) ].
\label{eq:complete_forward_model}
\end{equation}

\textbf{Source separation algorithm: } Given the acquisition model, our MCA algorithm is performed by solving the following convex optimization problem
\begin{subequations}
\label{eq:convex_relaxed_objective-final_formulation}
\begin{align}
\label{eq:objective}
\argmin_{\bs L, \bs x_d, \bs x_p} & \quad  \frac{\delta}{2} \lVert \bs Y - \bs L  - \varphi * R[\bs p (\bs x_d + \bs x_p)^\top ] \rVert_\delta^H,  \\
\label{eq:rank-constraint}
\text{s.t.} &\quad  \bs L \in \text{span} ( \bs U_r^{*}), \\
\label{eq:disk-constraint}
&\quad \lVert \bs \Psi^\top \bs x_d\rVert_1 \leq \tau_d,\\
\label{eq:planet-constraint}
&\quad \lVert\bs x_p\rVert_1 \leq \tau_p \,,\\
\label{eq:positivity-constraint}
& \quad \bs L,  \bs x_d,\bs x_p \geq 0.
\end{align}
\end{subequations} 
Each elements of the problem~\eqref{eq:convex_relaxed_objective-final_formulation} is justified by physical priors which are listed below.

\textbf{Fidelity term~\eqref{eq:objective}:} As shown in~\cite{pairet2019stim}, the speckle noise follows a sub-exponential distribution which implies $P\left (\bs N_s > \epsilon  \right) \leq\exp \left( -c \min( \epsilon^2/\delta^2, \epsilon/\delta) \right)$. This indicates that the negative log-likelihood of the speckle is approached by the Huber-loss function~\citep{huber1992robust} defined as 
 \begin{align}
| a |_\delta = 
\begin{cases} 
\frac{1}{2\delta} a^2 &\text{if } |a| \leq \delta \\
| a | - \frac{\delta}{2} &\text{if } |a| > \delta,
\end{cases}
\end{align}
which is well-suited for the fidelity term. The metric $ \lVert \cdot \rVert_\delta^H$ is simply defined as $ \lVert \bs a \rVert_\delta^H = \sum_i  |a_i|_\delta$, for a given vector $\bs a$. 

\textbf{Static part of the speckles~\eqref{eq:rank-constraint}:} As it is common in the literature of background-foreground separation~\citep[see, \eg][]{zhou2011godec}, we use a low-rank structure to model the static part of the speckle field, \eg impose $\lVert \bs L \rVert_*\leq \tau_L$ . However, the large intensity discrepancy between the disk and the star makes the use of the nuclear norm impractical. Indeed, in a dummy low-rank plus sparse problem~\cite{pairet2020mayonnaise}, we observed that, as the intensity of the low-rank component increases, the quality of the estimation of the sparse component becomes overly sensitive to the accuracy in the estimation of $\tau_L$. For ADI datasets, the intensity of $\Lmodel$ is typically around $10^3$ larger than the intensity of $\Xmodel$. In this case, even when $\tau_L$ is only a percent above or below the groundtruth, the recovery of $\bs x$ is unsuccessful.
Fortunately, the greedy algorithm presented in~\cite{Pairet2018reference} yields a decent estimate $\bs L^*$ of $\Lmodel$. Hence inspired by the work of~\cite{eftekhari2018weighted}, we relax the low-rankness imposed on $\Lmodel$ by forcing it to lie in the span of the first $r$ columns of $\bs U^*$, where $\bs U^*$ is given by the SVD decomposition of $\bs L^*$, as shown in~\eqref{eq:rank-constraint}.

\textbf{Spatial structure of the disks~\eqref{eq:disk-constraint}:} the disk is regularized by promoting its sparsity in the shearlets domain, as they have been shown to be efficient for sparsely representing multivariate data containing edges and curved structures~\cite{kutyniok2012introduction}.

\textbf{Include exoplanetary signal~\eqref{eq:planet-constraint}:} Following the premise of the MCA algorithm, we leverage the morphological diversity between the disk and the exoplanets to separate these two sources $\bs x = \bs x_d + \bs x_p$. After deconvolution, the exoplanetary signal $\bs x_p$ has an optimally sparse representation in the direct domain, whereas its shearlets representation involves more terms.

\textbf{Positivity of the images~\eqref{eq:positivity-constraint}:} since the signals of interest are positive, we finally enforce that  $\bs L, \bs x_d, \bs x_p \geq 0$.

\begin{figure}
  \centering
  {\includegraphics[width=0.2\textwidth]{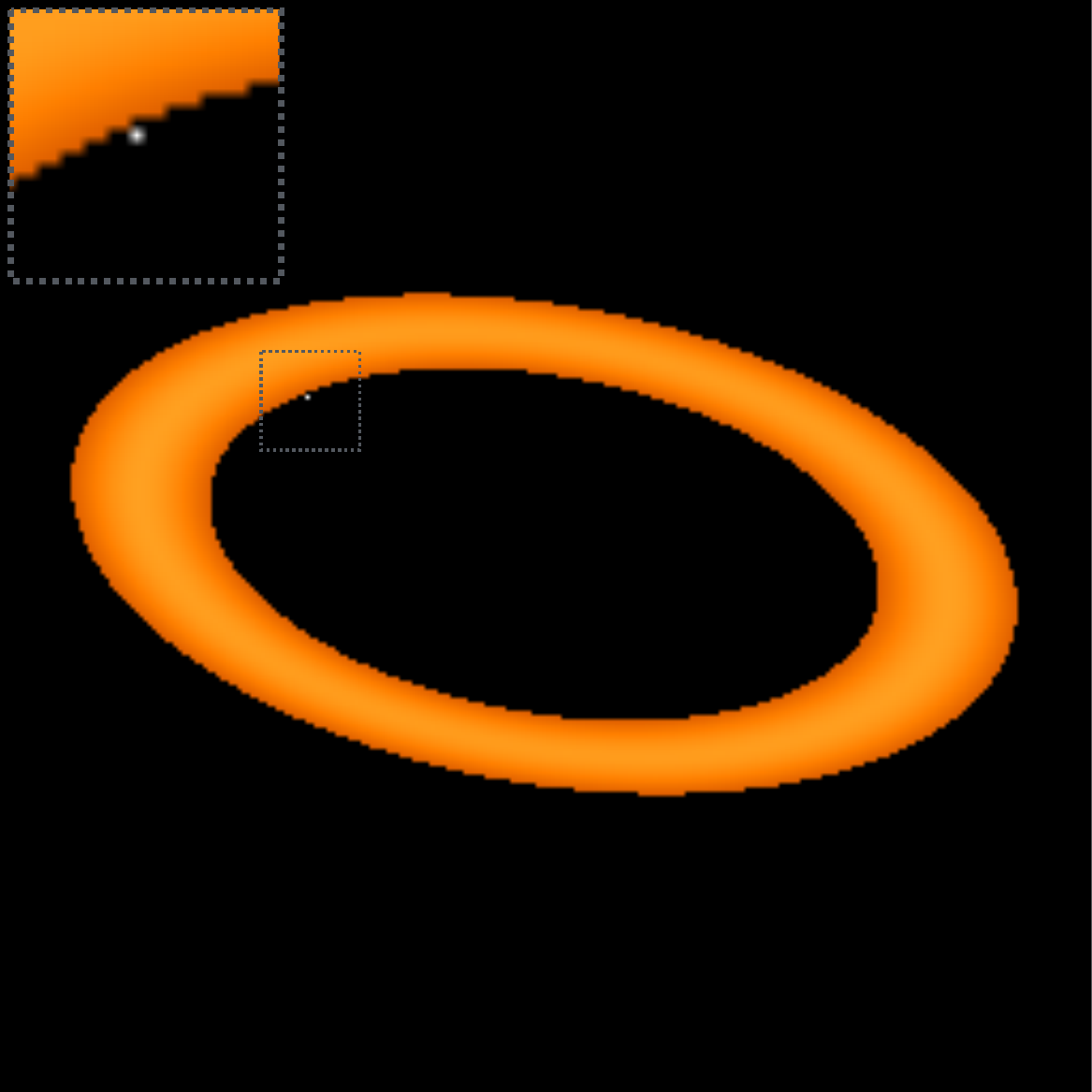}}
  {\includegraphics[width=0.2\textwidth]{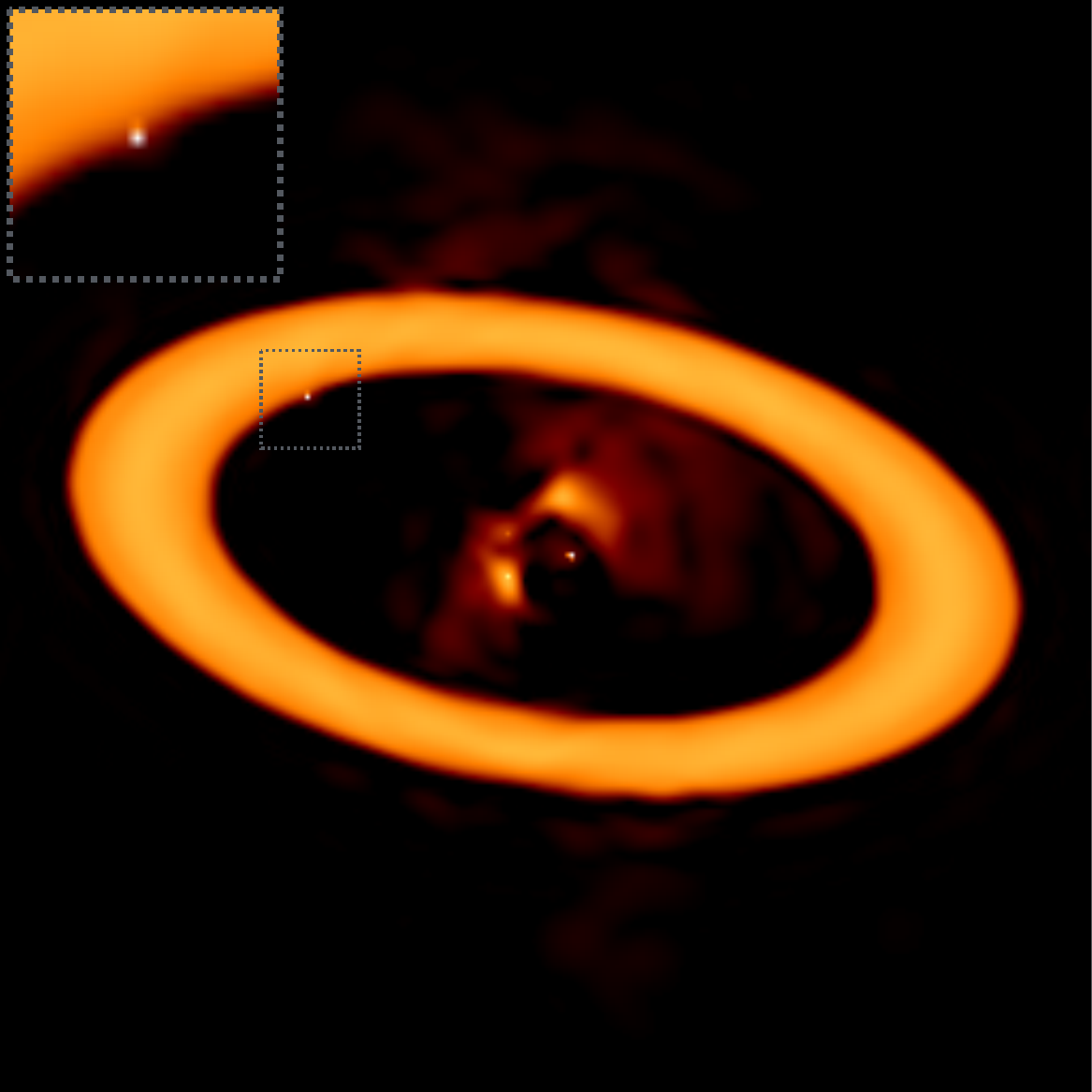}}
\caption{Left: disk and exoplanet injected in an empty ADI cube with a contrast of $10^{-5}$. Right: the recovery with our MCA approach. The shape of the disk is preserved and the exoplanetary signal is recovered.}
\label{fig:disk_planet}
\sq
\sq
\end{figure}

\section{Numerical experiments: }
\sq
We solve~\eqref{eq:convex_relaxed_objective-final_formulation} with the Primal-Dual Three-Operator splitting~\citep[PD3O][]{yan2018new}. The gradient of~\eqref{eq:objective} requires the computation of $R^\top$, which is done using the autograd functionality of \textsc{PyTorch}~\citep{paszke2017automatic} along with \textsc{kornia}, a \textsc{PyTorch}-based computer vision toolbox~\citep{riba2020kornia}. For the shearlets transform, we use the \textsc{python} version of the \textsc{ShearLab 3D} toolbox~\citep{kutyniok2016shearlab}). Fig.~\ref{fig:HR4796A} (right) shows the HR~4796A image obtained our method. We can see that the recovered throughput is about 8 times that of PCA. The overall shape of the disk is also more physically sound, \eg there is no negative intensities and the shape is closer to estimations obtained with physical models (see for instance~\cite{milli2017near}).

To illustrate the capability of our algorithm to faithfully recover both a disk and a exoplanet, we used the \textsc{VIP} toolbox~\cite{2016ascl.soft03003G} to create a disk $\bar{\bs x}_d$ with a exoplanet $\bar{\bs x}_p$ displayed in Fig.~\ref{fig:disk_planet} (left). Both the $\bar{\bs x}_d$ and $\bar{\bs x}_p$ were convolved with the telescope PSF before being injected in an empty ADI cube. We used our approach to recover $\bar{\bs x}_d$ and $\bar{\bs x}_p$, the result is shown in Fig.~\ref{fig:disk_planet} (right). We can see that our method was able to reproduce the shape of the disk faithfully and to recover the exoplanetary signal. 

\sq
\section{Conclusion}
\sq

We presented a MCA framework to image circumstellar disks and exoplanets from ADI datasets. Our method leverages physical knowledge to produce faithful images. To best of our knowledge, our method is the first to include MCA and deconvolution in ADI post-processing, allowing to disentangle exoplanets from circumstellar disks. Although our method is able to recover on-sky signals from ADI data, Fig.~\ref{fig:disk_planet} (right) also features intensities in the center that do not correspond to the injected signal. Future work should include a procedure to asses the quality of the output, as for instance, adapted hypothesis testing.

\clearpage

\small
\bibliography{biblio}{}
\bibliographystyle{plain}

\end{document}